\documentclass[sigconf]{aamas}

\usepackage{balance} 

\usepackage{cleveref}
\usepackage{siunitx}

\renewcommand\footnotetextcopyrightpermission[1]{} 
\setcopyright{ifaamas}
\acmConference[AAMAS '24]{Proc.\@ of the 23rd International Conference
on Autonomous Agents and Multiagent Systems (AAMAS 2024)}{May 6 -- 10, 2024}
{Auckland, New Zealand}{N.~Alechina, V.~Dignum, M.~Dastani, J.S.~Sichman (eds.)}
\copyrightyear{2024}
\acmYear{2024}
\acmDOI{}
\acmPrice{}
\acmISBN{}

\acmSubmissionID{<<EasyChair submission id>>}

\title{Preliminary Study of the Impact of AI-Based Interventions on Health and Behavioral Outcomes in Maternal Health Programs}

\author{Arpan Dasgupta}
\affiliation{
  \institution{Google Research India}
  \city{Bengaluru}
  \country{India}}
\email{arpandg@google.com}

\author{Niclas Boehmer}
\affiliation{
  \institution{Harvard University}
  \city{Boston}
  \country{USA}}
\email{nboehmer@g.harvard.edu}

\author{Neha Madhiwalla}
\affiliation{
  \institution{ARMMAN}
  \city{Mumbai}
  \country{India}}
\email{neha@armman.org}

\author{Aparna Hedge}
\affiliation{
  \institution{ARMMAN}
  \city{Mumbai}
  \country{India}}
\email{aparnahegde@armman.org}

\author{Bryan Wilder}
\affiliation{
  \institution{Carnegie Mellon University}
  \city{Pittsburgh}
  \country{USA}}
\email{bwilder@andrew.cmu.edu}

\author{Milind Tambe}
\affiliation{
  \institution{Harvard University, Google Research India}
  \city{Boston}
  \country{USA}}
\email{milindtambe@google.com}

\author{Aparna Taneja}
\affiliation{
  \institution{Google Research India}
  \city{Bengaluru}
  \country{India}}
\email{aparnataneja@google.com}

\begin{abstract}
    Automated voice calls are an effective method of delivering maternal and child health information to mothers in underserved communities. One method to fight dwindling listenership is through an intervention in which health workers make live service calls. Previous work has shown that we can use AI to identify beneficiaries whose listenership gets the greatest boost from an intervention. It has also been demonstrated that listening to the automated voice calls consistently leads to improved health outcomes for the beneficiaries of the program. These two observations combined suggest the positive effect of AI-based intervention scheduling on behavioral and health outcomes. This study analyzes the relationship between the two. Specifically, we are interested in mothers' health knowledge in the post-natal period, measured through survey questions. We present evidence that 
    improved listenership through AI-scheduled interventions leads to a better understanding of key health issues during pregnancy and infancy. This improved understanding has the potential to benefit the health outcomes of mothers and their babies. 
\end{abstract}


\newcommand{\BibTeX}{\rm B\kern-.05em{\sc i\kern-.025em b}\kern-.08em\TeX}

\usepackage{todonotes}

\begin{document}


\pagestyle{plain}
\fancyhead{}


\maketitle 


\section{Introduction}

Mobile Health (mHealth) programs can make essential health information accessible to less privileged communities. These programs utilize the large accessibility of mobile phones to spread critical health information but often suffer from beneficiary's loss of interest and subsequent drops in listenership over time. To address this, interventions such as a call or a visit from a community health worker can be an effective tool that keeps beneficiaries engaged in the program. However, the question of who should receive an intervention is a non-trivial prediction and planning problem. Previous work has established that AI can be used to schedule such interventions effectively in some mHealth programs, resulting in a significant increase in engagement in the program~\cite{verma2023deployed}. 

We partner with ARMMAN \cite{ArmmanFoundation}, an India-based non-profit organization that offers mHealth programs to increase awareness of antenatal and postnatal health amongst mothers. We focus on their mMitra \cite{mMitra} program, which is the second-largest maternal mHealth program in the world.
In this program, weekly automated voice messages deliver essential maternal health information to the beneficiaries. There is a limited number of live service calls that can be conducted by health workers every week to boost beneficiaries' engagement due to limited support staff in the NGO.
SAHELI \cite{verma2023deployed}, a project developed in the context of mMitra, is the first-ever large-scale deployment of AI in a mHealth program that effectively allocates these limited intervention resources. 
We study the effects of AI-scheduled interventions on the knowledge and behavior of the beneficiaries enrolled in mMitra through a conducted survey.

Previous studies \cite{hegde2016assessing,murthy2020effects} established that mothers who consistently listen to mMitra's automated voice messages have improved infant care practices and knowledge of maternal practices. In addition, previous studies \cite{verma2023deployed, boehmer2024evaluating,DBLP:conf/aaai/MateMTMVSHVT22} have also shown that AI-scheduled interventions boost engagement in the program, as they increase the amount of time the beneficiaries listen to the automated voice calls in a statistically significant. 
As a result, AI-scheduled interventions lead to an increased exposure of beneficiaries to critical health information. In contrast, if interventions are scheduled uniformly at random, studies were unable to establish a significant effect \cite{DBLP:conf/aaai/MateMTMVSHVT22}. 
This suggests an intuitive correlation between AI-scheduled interventions and improved health practices. However, no previous work has linked the usage of AI assistance and improved health outcomes.

In this work, we aim to establish a correlation between AI-scheduled interventions and improved behavioral and health outcomes in the mMitra program. For this, we conduct a survey on the beneficiaries, which aims to assess the beneficiaries' knowledge of good health practices discussed in the automated voice calls. We hypothesize that the increase in listenership resulting from the AI-scheduled interventions should lead to improved knowledge and thus better survey outcomes for people receiving an intervention. To test this hypothesis, we conducted a randomized control trial with two arms of beneficiaries with one \emph{intervention arm} receiving interventions as scheduled by the AI and a second \emph{control arm} receiving no interventions. We compare the knowledge and behavioral outcomes of beneficiaries in the two arms using the results of the conducted survey.




In our analysis and evaluation of the study, we establish a statistically significant increase in listenership caused by the AI-scheduled interventions. Moreover, we also observe generally better behavior and health practices in the intervention arm (as measured by the survey). However, we are unable to establish a statistically significant difference between the two arms with a high level of certainty due to small sample sizes and large amounts of noise in the responses. Our result calls for a modified larger-scale study to better establish our hypothesis, which we are running at the moment.


In sum, in this work, we showcase preliminary results demonstrating the impact of AI-scheduled interventions on the health and behavior of mothers enrolled in mMitra, a maternal health program. We also establish a statistically significant increase in listenership in the program.
The paper is organized as follows. In \Cref{sec:sec_prev_work}, we briefly discuss previous work in this domain. In \Cref{sec:sec_setup}, we discuss the setup of the study including intervention scheduling and the conducted survey. In \Cref{sec:sec_results}, we describe our analysis method for the survey and our obtained results. 

\section{Related Work} \label{sec:sec_prev_work}

The problem of how to allocate limited resources comes up in several domains which require planning.
Restless multi-arm bandits (RMABs) are a popular tool for such sequential allocation problems in uncertain environments. In particular, RMABs have shown to be very useful in applications such as anti-poaching patrols \cite{qian2016restless}, multi-channel communication \cite{liu2010indexability}, scheduling \cite{bagheri2015restless, yu2018deadline}, aerial vehicle routing \cite{zhao2008myopic}, and machine maintenance and sensor monitoring \cite{glazebrook2006some}. These limited resource allocation problems naturally also appear when planning interventions in mHealth programs \cite{mate2020collapsing}.

Past work has established that the health information provided in mHealth programs leads to improved infant care practices and knowledge of maternal practices among mothers \cite{hegde2016assessing,murthy2020effects}. In particular, \citet{hegde2016assessing} use a randomized controlled trial to measure the effect of tailored voice calls on mothers in mMitra.  \citet{hegde2016assessing} establish statistically significant results for improved infant care knowledge among mothers as well as a direct impact on infant health as measured by their birth weight.

These results motivate ARMMAN to try to boost beneficiary's listenership through service calls by health workers. 

In collaboration with ARMMAN, \citet{DBLP:conf/aaai/MateMTMVSHVT22} describe an AI-based method for scheduling intervention calls. This method decides how service calls are allocated using the RMAB framework, where each beneficiary is modeled as a Markov decision process. Their method was initially tested in simulations, and subsequently in a field study before it was finally deployed at scale in practice \cite{verma2023deployed}. A fundamental challenge in SAHELI has been to learn the transition probabilities of the Markov decision processes modeling beneficiaries. After multiple refinement steps, \citet{51909} and \citet{DBLP:conf/atal/VermaM0MHTT23} utilized decision-focused learning (DFL) \cite{shah2022decision} for RMABs to learn transition probabilities as to improve the performance of the program in deployment.

So far, the observable objective that is optimized by SAHELI and other intervention scheduling programs for ARMMAN is the mother's listenership of automated voice calls and hence program's performance is always measured in terms of improvement in listenership metrics.  However, no correlation has been shown between AI-scheduled interventions and behavioral outcomes, a gap that we investigate in our study.

\section{Setup of the Study}\label{sec:sec_setup}

As the first step in the study, we divided the registered beneficiaries into cohorts based on their time of enrollment. Subsequently, for each cohort, we divide the beneficiaries into intervention and control groups. The automated voice calls containing health information are received by everyone in both arms throughout their enrollment in the program. Those enrolled in the intervention group are eligible for receiving interventions. The AI algorithm decides which beneficiaries receive interventions in the form of live service calls from health workers in a given week. In the end, a survey is conducted on subsets from both intervention and control groups to measure the behavioral and health knowledge of the beneficiaries.

\subsection{Experiment Arms}

\subsubsection{Cohorts}
The study was conducted in three cohorts with a combined number of $60464$ beneficiaries.
\begin{itemize}
    \item Cohort 1: 27688 beneficiaries. Registered between 15th August 2022 to 31st of September 2022.
    \item Cohort 2: 13972 beneficiaries. Registered between 1st October 2022 to 31st October 2022.
    \item Cohort 3: 18804 beneficiaries. Registered between 1st November 2022 to 31st November 2022.
\end{itemize}
As explained in \Cref{subsub:int}, these cohorts were not viewed as fully independent by the program and are instead primarily used to determine when beneficiaries are eligible for receiving an intervention. 

\subsubsection{Division Into Arms}

For each cohort, we split the beneficiaries into intervention and control arms so that attribute distributions between arms are similar. This resembles covariate adaptive randomization \cite{lachin1988randomization} where distributions of the covariates are balanced across the groups. We balance between the following attributes:

\begin{description}
    \item[Engagement states] \phantom{a}
    \begin{itemize}
        \item For every beneficiary and a specific automated voice call, we define its engagement state E@T at threshold T as E@T = 1 if the beneficiary listened to at least T seconds of the call and 0 otherwise. 
        \item For every beneficiary, we calculate E@T\_w for $w$ weeks before the expected intervention start date for the cohort.
        \item We strive for achieving approximately equal values of E@T\_w for $T\in \{1, 5, 10, 30, 100\}$ and $w\in \{1,2,3\}$ between arms to ensure a similar distribution of listenership profiles across the two arms.
    \end{itemize}
    \item[Demographic Features]
    We consider the gestational age of beneficiaries in terms of their current trimester as a feature for preparing cohorts. This quantity is calculated by dividing gestational age in weeks by $14$ which gives 4 bins of equal size. We make sure that both groups have an equal number of beneficiaries in each of the trimesters. This also makes sure that both groups contain an equal number of beneficiaries who have already delivered.
\end{description} 
To ensure a balanced distribution of these attributes between arms we first create a vector $Y$ for each beneficiary by appending the attributes. We then split the beneficiaries into two equal groups using $Y$ as the splitting factor \cite{51909}. This is done by treating $Y$ as a class label and using a stratified splitter to split it into two equal groups. We use the stratified option in \textit{test\_train\_split} in the sklearn \cite{pedregosa2011scikit} package. Since we have enough beneficiaries in the cohort, we are able to get an exact split.

\subsection{Conducting Interventions}
Interventions in mMitra are service calls made by healthcare workers that aim to boost the future listenership of automated messages of the called beneficiary.

\subsubsection{Number of Interventions per week.}\label{subsub:int}
Interventions began on 21st November 2022. We only intervene on beneficiaries that have been present for at least $6$ weeks in the program. In the beginning, we are only allowed to act on beneficiaries from cohort $1$, then after some time on beneficiaries from cohorts $1$ and $2$, and then finally on beneficiaries from all three cohorts. 
This was done to simulate the deployment which considers several months of enrollments at the same time:

\begin{itemize}
    \item 21st November 2022 to 12th December 2022 (4 weeks) - consider only Cohort 1 for interventions.
    \item 19th December 2022 to 9th January 2022 (4 weeks) consider Cohort 1 + Cohort 2 for interventions.
    \item 16th January 2022 to 13th February 2022 (5 weeks) consider Cohort 1 + Cohort 2 + Cohort 3 for interventions.
\end{itemize}
We conduct approximately $1000$ interventions per week while ensuring that each beneficiary can be intervened on only once. We end up conducting interventions on about $12000$ beneficiaries which accounts for about $43\%$ of the intervention arm.

\subsubsection{Eligibility for Interventions.}
Beneficiaries are eligible for interventions  under the following conditions. 
\begin{enumerate}
    \item Active status - they are still enrolled in the program and receive automated voice messages.
    \item They picked up at least 1 voice message in the last 4 weeks before the start of the intervention period for the respective cohort.
    \item Repeat Intervention - beneficiaries have not received a previous service call.
\end{enumerate}

\subsubsection{Conducting Interventions.}

Each week, the DFL-RMAB \cite{51909} algorithm  which is our AI algorithm of choice, determines the set of beneficiaries from the intervention arm who will receive an intervention. We collect all beneficiaries from the intervention arm that have received an intervention in some week in a list $I_D$. We also simulate the AI algorithm on the control arm to determine the set of beneficiaries that would have been selected for an intervention  (assuming we conducted the same number of interventions as in the intervention arm). As in the intervention arm, beneficiaries in the control arm cannot be selected multiple times. We collect the beneficiaries from the control arm that have been selected by the algorithm in some week into a list $I_C$. 
We create an intervention list $I$ that combines $I_D$ and $I_C$. 
The idea is that we later compare the behavior of beneficiaries from $I_D$ and $I_C$, as we can think of beneficiaries from $I_C$ as the counterparts of those from $I_D$.

\subsection{Health Survey}

\subsubsection{Conducting the Survey}

The health survey is conducted on the beneficiaries from the intervention list~$I$ between September and November in $2023$. Since the program is oriented towards helping beneficiaries who lack resources the most, we perform the survey only on beneficiaries who are ``high-risk''. A beneficiary is considered ``high-risk'' if they satisfy at least one of the following three conditions:  low level of formal education (less than grade $10$), low-income family (less than $10000$ INR per month) or they do not own the phone themselves. Around $50\%$ beneficiaries enrolled in the program are high-risk. 

\begin{figure*}[h]
  \centering
  \includegraphics[width=0.8\linewidth]{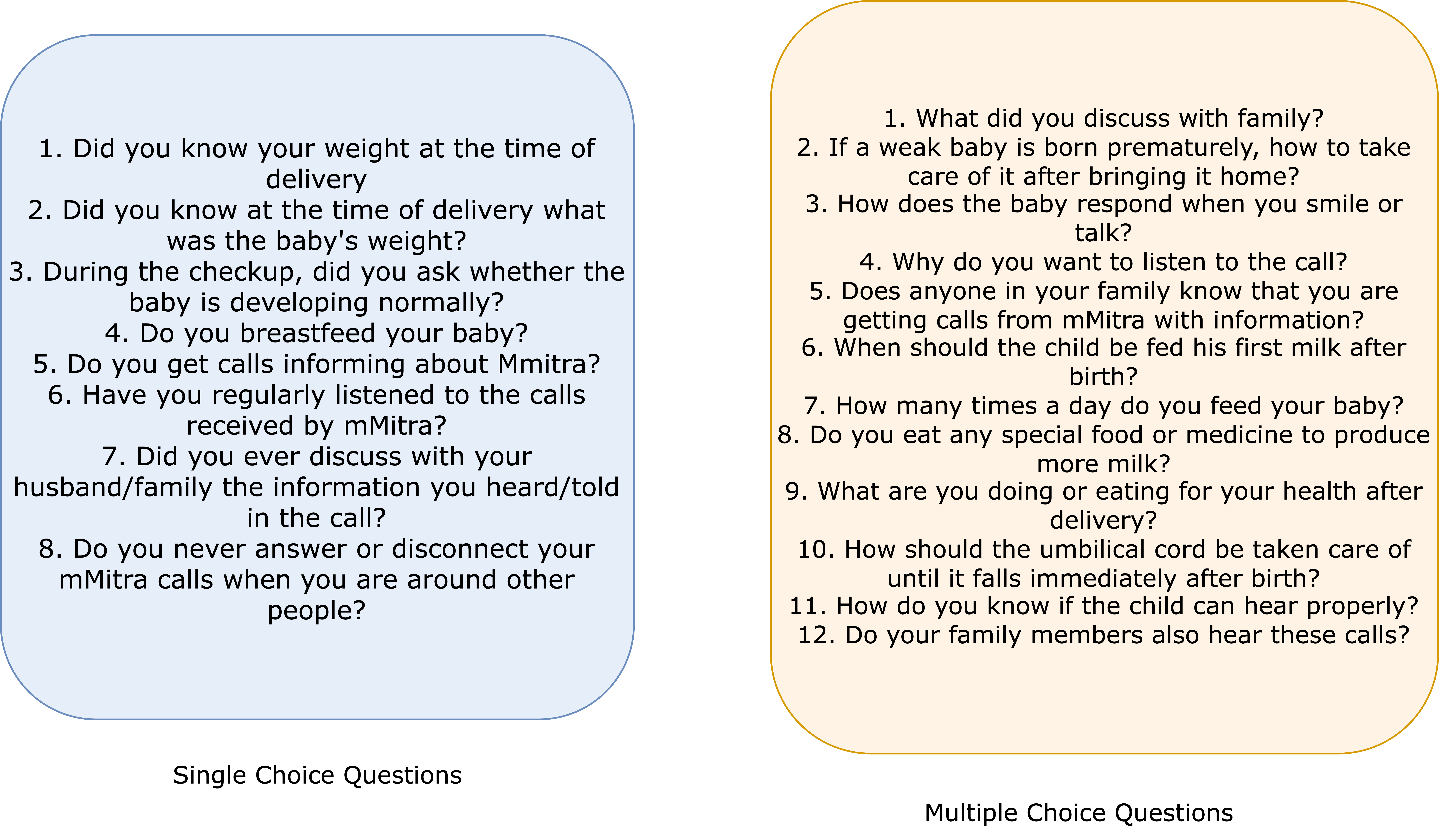}
  \caption{Questions asked in the survey.}
  \label{fig:questions}
\end{figure*}

The subset of the "high-risk" women who give birth between the intervention and survey call and have been in the program for at least 3 months are called by a health worker and asked to answer the questions from the survey. However, the survey calls are only picked up by a fraction of beneficiaries ($3376$ out of the $6448$ called). This provides a challenge for the evaluation of our study, as we only have access to the "outcome" of a subset of beneficiaries, i.e., those who were willing to answer to the survey questions (in particular, this group of beneficiaries is not chosen uniformly at random).  This makes it for instance necessary to re-balance the control and intervention group for the final comparison.

\subsubsection{Survey Questions.}

Each participant was asked $20$ questions with the intent of measuring their engagement with the program and measuring knowledge in different areas such as health practices. This evaluation is guided by the content of the automated voice messages and assesses how well the beneficiary received the messages. Concretely, the categories covered in the survey are: engagement with the program, knowledge, breastfeeding practices, communication, and health supplements. \Cref{sec:appendix}  contains a description of  the questions in each category. \Cref{fig:questions} shows the list of questions. 
The beneficiaries obtain a score for their answer in each question.

Out of the $20$ questions, $8$ correspond to single choice (Yes/No) questions. The remaining $12$ correspond to questions where scores are calculated based on multiple possible correct answers with some answers potentially contributing a higher score than others. Beneficiaries may choose one or more of these answers and their score is the total score for correctly identified answers in the question.

\section{Results and Analysis} \label{sec:sec_results}

\subsection{Comparison Methodology}

We compare the survey responses of beneficiaries in the  intervention group and the control group. However, in doing so, we face multiple challenges. Firstly, not everyone from the list $I$ picks up and answers the survey call. Secondly, some beneficiaries from the intervention list $I_D$ of the intervention arm do not pick up the intervention call. 
To ensure that the beneficiaries from the intervention and control arms that we compare to each other are similar in distribution, we use a matching method to pair beneficiaries from different arms.

\subsubsection{Beneficiary Matching}

We perform a one-to-one pairing of beneficiaries that answered the survey in the two arms by using matching methods with feature variables \cite{rubin1979using,rosenbaum1985constructing}. The features used for this matching are the average listenership over the last $6$ weeks from the intended date of intervention, gestational age, and the number of children they conceived previously. Let $X_i$ be the feature vector of the $i_\text{th}$ beneficiary and $X$ the feature matrix consisting of stacked vectors of all beneficiaries who responded to the survey.
We define a closeness metric \cite{stuart2010matching} between the feature vectors of two beneficiaries as their Mahalonobis distance
\begin{equation}
    D_{ij} = (X_i - X_j)' \Sigma^{-1} (X_i - X_j),
\end{equation}
where $\Sigma$ is the variance matrix of $X$ in the pooled dataset.

Finally, for each beneficiary in the intervention group who responded to the survey and picked up the intervention, we greedily pick the closest beneficiary from the control group who responded to the survey using the above-mentioned distance metric. Once a control beneficiary is matched, we no longer consider it for further matching.
Finally, we keep the beneficiaries that are part of one pair to obtain two sets of beneficiaries $S_D\subseteq I_D$ and $S_C\subseteq I_C$ for the intervention and control arms, respectively. 

\subsubsection{Improving Beneficiaries}

As we will discuss in \Cref{sub:analys}, establishing effect sizes in the full cohort of beneficiaries is challenging, among others, because some beneficiaries do not show an improvement in listenership after the intervention call. 
Recalling that previous work \cite{hegde2016assessing} has identified that listening to automated calls leads to better health outcomes, we expect that beneficiaries who experienced the greatest listenership boost through the intervention will show the most significant improvements in survey results.

\begin{figure*}[h]
  \centering
  \includegraphics[width=0.7\linewidth]{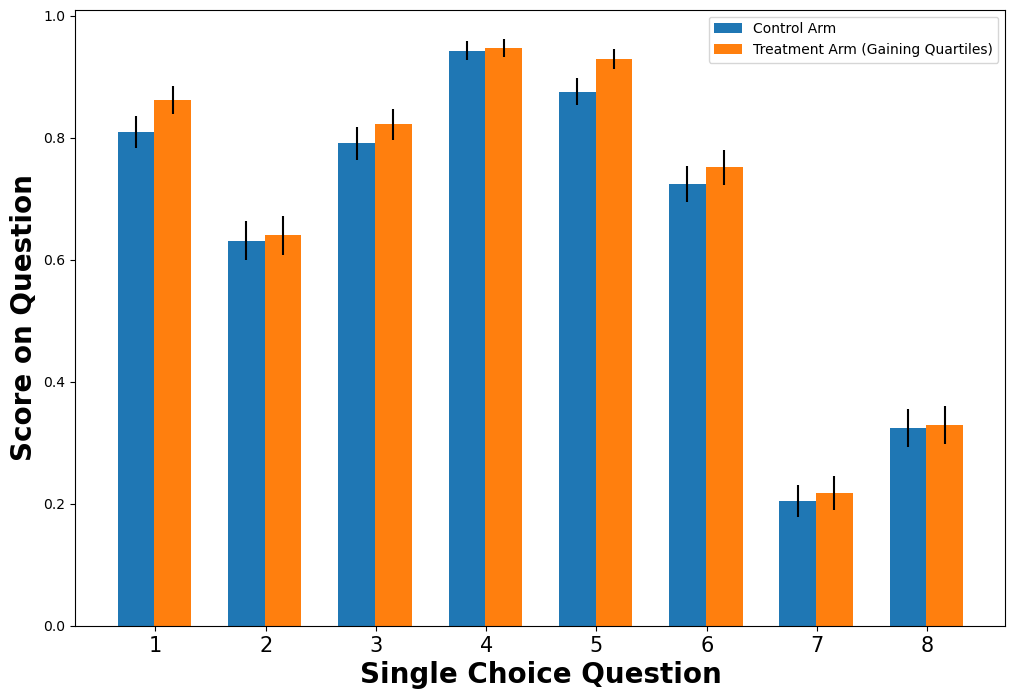}
  \caption{Score on single choice questions for beneficiaries who gained quartiles in listenership post-interventions and their corresponding beneficiaries from the control arm. The subset of beneficiaries from the intervention arm in all questions. The error bars represent the standard error in the measurement.
  }
  \label{fig:results_sc_1}
\end{figure*}

To this end, we formulate a method for identifying which beneficiaries have gained the most from the intervention. We define two listenership values for each beneficiary in the intervention list $I$, the average listenership over the past $6$ weeks before the scheduled intervention date, called the pre-listenership and the average listenership over the next $12$ weeks after the scheduled intervention date, called the post-listenership. The scheduled intervention date refers to the week in which the beneficiary from $I$ has been selected by the algorithm independent of whether they are in the treatment or control arm (and thereby independent of whether they actually received the intervention). Taking $12$ weeks of post-listenership (instead of $6$ weeks) allows us to measure the long-term gains from the interventions.

We calculate the quartile each beneficiary belongs to when compared with the other beneficiaries in $I$ for the pre- and post-listenership. For beneficiary $i$, we denote these values as $q_{1i}$ and $q_{2i}$, respectively.
We say that a beneficiary \emph{gains in quartiles} if $q_{2i} > q_{1i}$. 
In our analysis, we focus on the beneficiaries that gain in listenership, i.e., for which $q_{2i} > q_{1i}$. Specifically, in our analysis, we restrict our attention to the beneficiaries from the set $S_d$ with $q_{2i} > q_{1i}$ (called $R_d$) and the beneficiaries from $S_c$ matched to these beneficiaries (called $R_c$). Our final comparison involves $218$ pairs (i.e., $436$ beneficiaries in total).

Our method of selecting beneficiaries who have gained in quartiles is naturally only an approximation for the beneficiaries who benefit the most from an intervention. 
Note that a larger subsection of beneficiaries who received an intervention gain quartiles ($218$) compared to those who lose quartiles ($119$). This implies that interventions oftentimes lead to a gain in quartiles.

\begin{figure*}[h]
  \centering
  \includegraphics[width=0.65\linewidth]{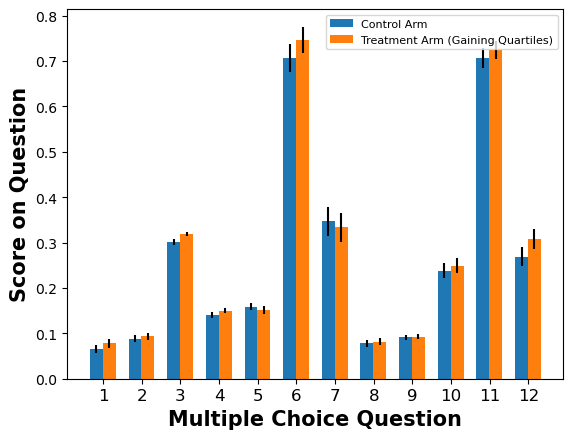}
  \caption{Score on multiple choice questions for beneficiaries who gained quartiles in listenership post-interventions and their corresponding beneficiaries from the control arm. Except on Question $5$, beneficiaries from the intervention arm perform on average better. 
  }
  \label{fig:results_mc_1}
\end{figure*}

\begin{figure*}[h]
  \centering
  \includegraphics[width=0.85\linewidth]{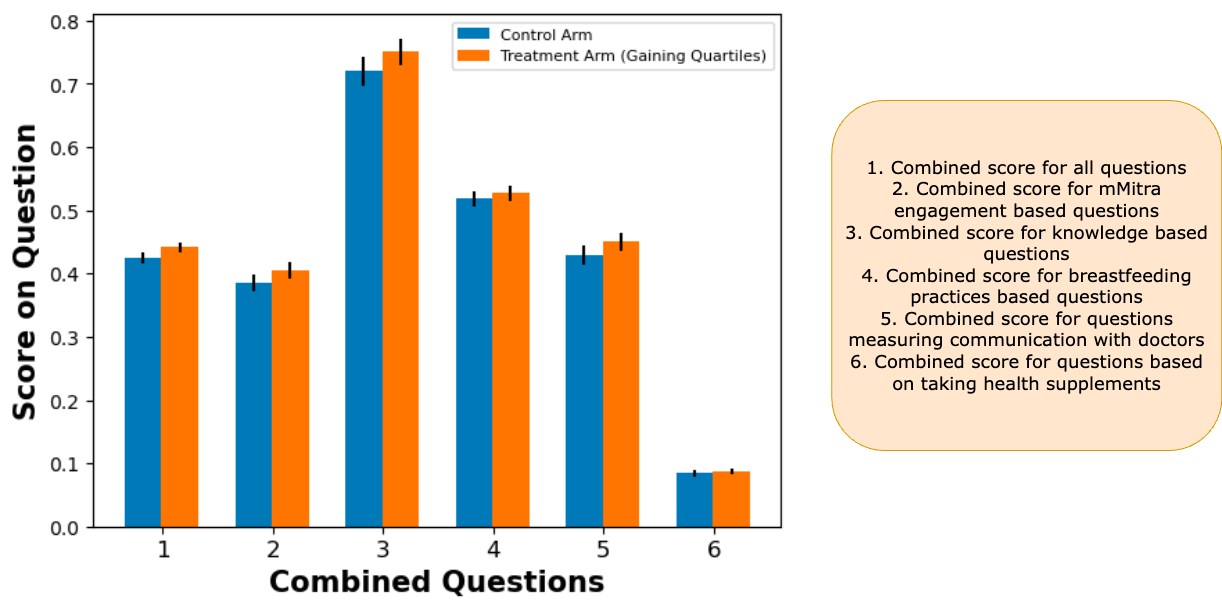}
  \caption{Score on groupings of questions for beneficiaries who gained quartiles in listenership and their corresponding beneficiaries from the control arm. Average scores in all categories are higher for beneficiaries from the intervention arm.}
  \label{fig:results_cc}
\end{figure*}

\subsection{Analysis}

\subsubsection{Establishing Improved Listenership}

We observe a consistent gain in listenership among beneficiaries who received interventions. In particular, following the work of  \citet{boehmer2024evaluating}, we compare (a) the beneficiaries from the intervention group that have been selected for an intervention (i.e., they are on $I_D$) and that responded to the survey call to (b) the beneficiaries from the control group that would have been selected for an intervention (i.e., they are on $I_C$) and responded to the survey call. 
We compare these groups using the subgroup estimator, see the work of \citet{boehmer2024evaluating} for details. We find that an intervention increases beneficiaries listenership compared to the control group by, on average, $7.43$ seconds per call over the 12 automated voice calls following the intervention, which leads to a summed additional listenership of $89.16$ seconds over the next $12$ calls. The $95\%$ confidence interval for this value is $[45.516,132.936]$ and the hypothesis that interventions have a positive non-zero effect on listenership can be accepted with a p-value of $\num{6.4656e-05}$. This allows us to conclude a statistically significant positive effect of interventions on beneficiary's listenership. Notably, the positive effect of interventions has already been established in previous studies \cite{DBLP:conf/aaai/MateMTMVSHVT22,DBLP:conf/atal/VermaM0MHTT23,boehmer2024evaluating}; however, our identified effect sizes and achieved confidence level are advantageous. 

\subsubsection{Establishing Behavioral and Health Benefits}\label{sub:analys}

We now analyze the beneficiary's performance in the survey. We will start by focusing on the beneficiaries on which interventions have been effective. 
\Cref{fig:results_sc_1,fig:results_mc_1} show the results on single choice and multiple choice questions, respectively. 
We plot the average score obtained on each question for the groups $R_d$ and $R_c$, i.e., beneficiaries from the treatment group whose listenership improved and who answered the survey and their counterparts. The empirical standard error is included as error bars in the plot to signify the possible error in the comparison.
Note that independent of the group membership beneficiaries achieve generally higher scores on the single choice questions (arguably, because it is easier to respond to Yes/No questions during a survey). 

The average score of the intervened beneficiaries from $R_d$ is higher than the average score of the control beneficiaries from $R_c$ on all but one  question.
For two questions, the difference between average scores is also statistically significant, i.e., single choice question $5$ and multiple choice question $3$. For the latter question ("How does the baby respond when you speak or talk?"), we can establish a difference with a p-value of $0.002$\footnote{Note that this value should be interpreted with a bit of caution, as we have some dependencies and confounding factors in our study, e.g., regarding the beneficiaries from which we received survey results.}, showing an improved awareness of the baby's behavior. 
A very high variance in answers is observed in some questions such as Question $6$ and $7$ from the multiple choice section, which ask about when a child should be first fed milk and how many times a day they should be fed. This trend could be due to either a lack of information among some mothers, or due to the noise in the way these questions were answered during the survey.

\Cref{fig:results_cc} plots the averaged summed scores on different groups  of questions.
In particular, the leftmost bars, depict the summed average score over all questions. 
We find that the difference between the two arms is larger than one standard error here and that we can establish a difference between the two arms with a p-value of $0.13$. While this is not enough to claim statistical significance, it is a strong hint towards the positive impact of the AI-scheduled interventions and motivates us to conduct a larger follow-up trial. 
\Cref{fig:results_cc} also covers the summed score in questions from different categories, that are, engagement with the program, knowledge, breastfeeding practices, communication, and health supplements. The questions comprising each category are provided in the appendix (\Cref{sec:appendix}). We notice that the engagement category provides the strongest trends and shows the largest difference between the arms. This is likely because the relationship between listenership and engagement with the program is a direct one, while the other outcomes interfere with more confounding factors.

In \Cref{fig:results_sc_3,fig:results_mc_3,fig:results_cc_3}, we show the results for all beneficiaries that filled out the survey from the intervention group and their counterpart from the control group, i.e., $S_D$ and $S_C$. Compared to the above analysis, we observe slightly weaker, yet still positive trends. In particular, the intervention group performs on average better in $7$ out of $8$ single choice questions and $8$ out of $12$ multiple choice questions. This leads to a better average performance across all question categories, yet the differences between the two groups are much smaller here than on the beneficiaries that gained quantiles. 
This is to be expected recalling that the gaining beneficiaries are the ones that respond most favorably to the intervention and will thus also be the ones with the highest knowledge gain.

\section{Conclusion and Future Work} \label{sec:sec_conc}
In conclusion, our study has shown that AI-scheduled interventions lead to significantly higher listenership in automated voice messages. 
Moreover, we provide some first evidence that these interventions also lead to improved behavioral and health outcomes as measured by our survey.
Together with the NGO, we aim to redesign the study with a more focused cohort and updated survey questions to better understand and correlate the impact of intervention-induced improved listenership on behavioral outcomes. 

\begin{acks}
We thank Suresh Chaudhary, Harshit Khaitan, and Shresth Verma for helpful discussions and for facilitating the study. 
\end{acks}



\newpage
\balance
\bibliographystyle{ACM-Reference-Format} 
\bibliography{main}


\newpage
\clearpage
\appendix



\section{Survey Appendix} \label{sec:appendix} 

\subsection{Questions and their answers} 

Single Choice Questions and expected answers:
\begin{itemize}
\item 0. Did you know your weight at the time of delivery : Yes
\item 1. Did you know at the time of delivery what was the baby's weight? : Yes
\item 2. During the checkup, did you ask whether the baby is developing normally? : Yes
\item 3. Do you breastfeed your baby? : Yes
\item 4. Do you get calls informing about Mmitra? : Yes
\item 5. Have you regularly listened to the calls received by mMitra? : Yes
\item 6. Did you ever discuss with your husband/family the information you heard/told in the call? : Yes
\item 7. Do you never answer or disconnect your mMitra calls when you are around other beneficiaries? : No
\end{itemize}


Multiple Choice Questions and expected answers with scores:
\begin{itemize}
    \item 0. What did you discuss with family? : ('weight', 0.33), ('cough', 0.33), ('Other Response', 0.33), ('baby is fussy', 0.33), ('constant watering', 0.33), ('breastfeeding problems', 0.33)
    \item 1. If a weak baby is born prematurely, how to take care of it after bringing it home? : ('entire black part of the breast', 0.33), ('pillow under the baby', 1), ('one breast to the other', 1), ('I hold the baby\'s head', 0.33)
    \item 2. How does the baby respond when you smile or talk? : ('Smiling back', 0.33), ('Watches us', 0.33), ('Shouts or speaks back', 0.34)
    \item 3. Why do you want to listen to the call? : ('regarding diet', 0.2), ('changes are happening', 0.2), ('answers to some questions', 0.2), ('information I don\'t get from doctors', 0.2), ('doing the right thing', 0.2)
    \item 4. Does anyone in your family know that you are getting calls from mMitra with information? : ('Husband', 0.25), ('Mother', 0.25), ('Relatives', 0.25), ('in-laws', 0.25)
    \item 5. When should the child be fed his first milk after birth? : ('an hour later', 1)
    \item 6. How many times a day do you feed your baby? : ('10-12 times', 1), ('every two hours', 1)
    \item 7. Do you eat any special food or medicine to produce more milk? : ('Traditional food', 0.05), ('Milk products', 0.15), ('Iron', 0.4), ('Calcium', 0.4)
    \item 8. What are you doing or eating for your health after delivery? : ('Traditional food', 0.083), ('Keeping ears/feet warm', 0.083), (‘boiling water and drinking hot water’, 0.083), ('Calcium tablets', 0.25), (‘Iron pills’, 0.25), ('Vitamin supplements', 0.25)
    \item 9. How should the umbilical cord be taken care of until it falls immediately after birth? : ('Must be kept clean and dry', 0.5), ('Nothing should be applied', 0.5)
    \item 10. How do you know if the child can hear properly? : ('Turns head in that direction when there is sound', 1.0), ('Looks at us when called', 0.5), ('Doctor can tell after checking up', 1.0)
    \item 11. Do your family members also hear these calls? : ('Sometimes', 0.5), ('Always', 1.0), ('Everyone in the family', 1.0), ('I tell them what the call was about', 0.25)

\end{itemize}

\subsection{Grouping of Questions}

The following is the grouping mechanism followed (SC and MC refer to single correct and multi correct respectively): 
\begin{itemize}
    \item Engagement : SC - [4, 5, 6, 7], MC -[3, 4, 11], 
    \item Knowledge : SC - [0, 1], 
    \item Breastfeeding Practices : SC - [3], MC - [5, 6, 7],
    \item Communication : SC - [2], MC - [0], 
    \item Supplements : MC - [7, 8]
\end{itemize}

\begin{figure*}[h]
  \centering
  \includegraphics[width=0.7\linewidth]{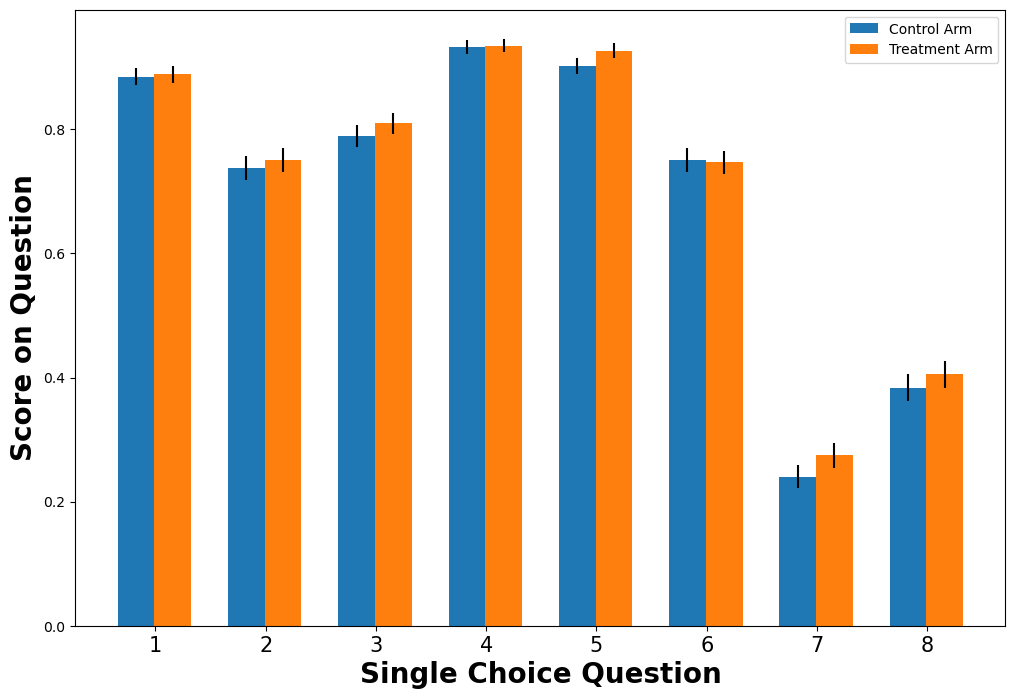}
  \caption{Average score on single choice questions for all beneficiaries.
  }
  \label{fig:results_sc_3}
\end{figure*}
\begin{figure*}[h]
  \centering
  \includegraphics[width=0.7\linewidth]{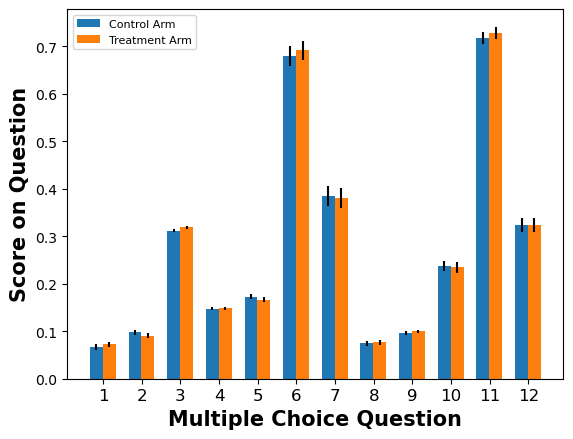}
  \caption{Score on multiple choice questions for all beneficiaries. 
  }
  \label{fig:results_mc_3}
\end{figure*}

\begin{figure*}[h]
  \centering
  \includegraphics[width=0.7\linewidth]{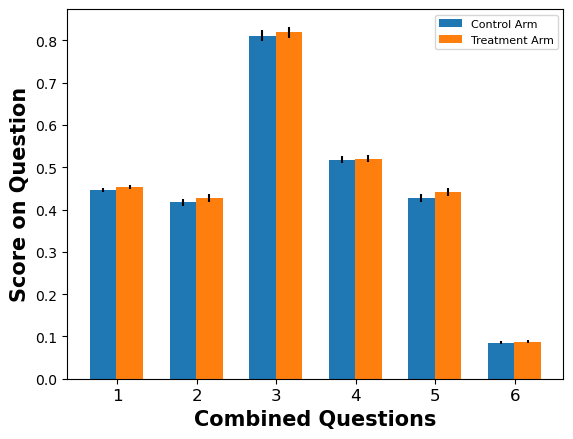}
  \caption{Score on groupings of questions for all beneficiaries.}
  \label{fig:results_cc_3}
\end{figure*}

\end{document}